\documentclass[twocolumn,showpacs,preprintnumbers,amsmath,amssymb,prl,superscriptaddress,aps,10pt]{revtex4-1}

\usepackage{graphicx}
\usepackage{color}

\begin{document}

\title{Full counting statistics of interacting quantum dots contacted by a normal metal and a superconductor}
\author{H. Soller}
\affiliation{Institut f\"ur Theoretische Physik, Ruprecht-Karls-Universit\"at Heidelberg,\\
 Philosophenweg 19, D-69120 Heidelberg, Germany}
\author{A. Komnik}
\affiliation{Institut f\"ur Theoretische Physik, Ruprecht-Karls-Universit\"at Heidelberg,\\
 Philosophenweg 19, D-69120 Heidelberg, Germany}

\date{\today}

\begin{abstract}
We investigate the effects of Coulomb interaction on charge transfer through a quantum dot attached to a normal and a superconducting lead. While for voltages much larger than the gap we recover the usual result for normal conductors,  for voltages much smaller than the gap superconducting correlations lead to a drastically different behavior. Especially, the usual charge doubling in the normal case is reflected in the occurence of quartets due to the onsite interaction.
\end{abstract}


\maketitle

\begin{figure}
\centering
\includegraphics[width=7cm]{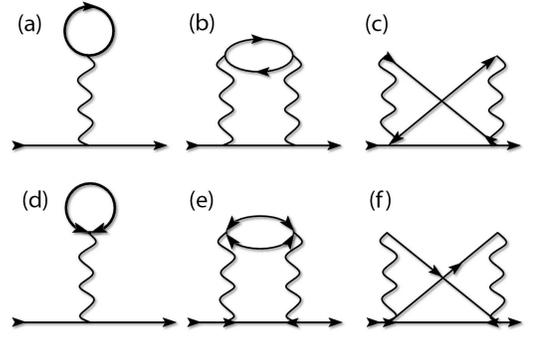}
\caption{Modified dot GFs: lines denote either normal or anomalous exact-in-tunneling GFs and the wiggly lines correspond to onsite interaction.}
\label{fig1}
\end{figure}

\begin{figure}[ht]
\centering
\includegraphics[width=7cm]{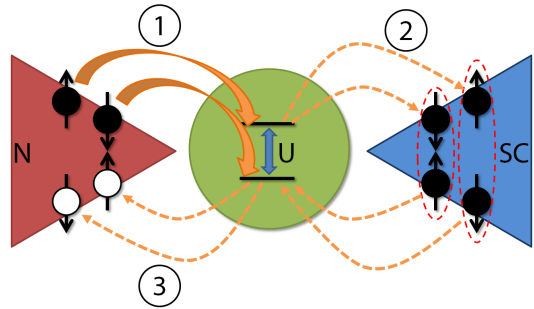}
\caption{Illustration of the quartic contribution to the CGF: two electrons hop onto the quantum dot (1) and `feel' their mutual repulsion $U$. This repulsion leads to correlated scattering onto the SC which results in two correlated ARs (2). The two holes are transferred further to the normal lead (3) which leads to a quartic contribution to the CGF.}
\label{fig33}
\end{figure}

\begin{figure}
\centering
\includegraphics[width=7cm]{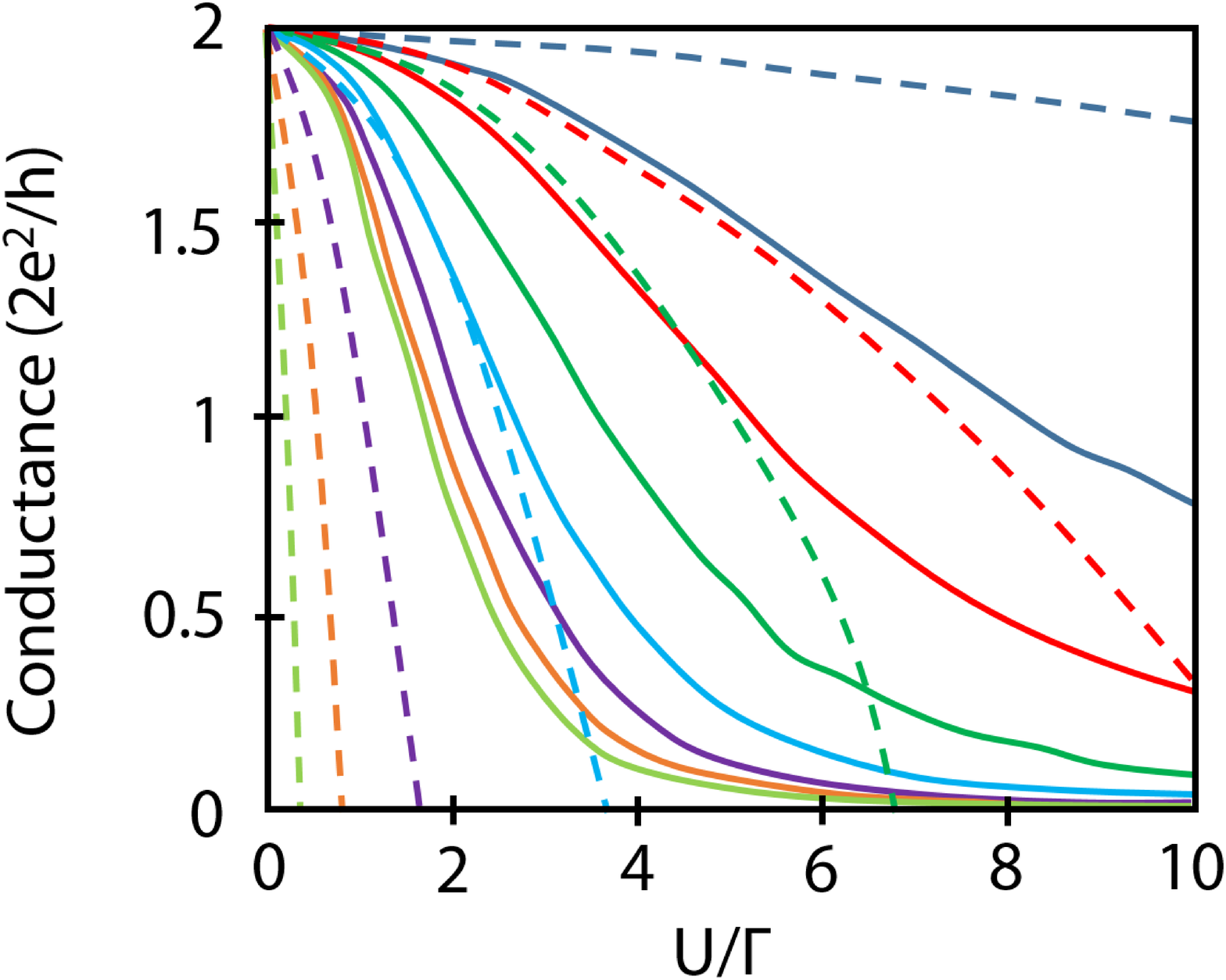}
\caption{Conductance at zero temperature as a function of $U/\Gamma$ and for different values of $\Gamma/\Delta$ taken from Eq. (\ref{condred}). From bottom to top $\Gamma/\Delta = 0.125, 0.25, 0.5, 1.0, 2.0, 4.0, 8.0$ shown as the dashed curves. The solid curves are the results from \cite{PhysRevB.63.094515}.}
\label{fig3}
\end{figure}

Hybrid nanoscale devices involving superconductors have recently attracted much attention both theoretically \cite{doi:10.1080/00018732.2011.624266,PhysRevLett.106.057001,PhysRevLett.106.257005} and experimentally \cite{Mourik25052012,heiblum,19829377} due to the possibility of generating Majorana fermions and using the superconductor as a source of spin-entangled electrons \cite{19829377}. These recent advances lead to the conclusion that nanoscale superconducting devices could be a future building block of quantum computers \cite{strunk}.\\
To exploit the correlations of electrons in superconductor hybrids a precise understanding of the effects of interactions, especially in quantum dots is necessary \cite{PhysRevLett.107.036801,PhysRevB.85.174512,PhysRevB.85.035419}. The correlations can be characterised by a number of quantities like the current-voltage characteristics, the current noise spectra, the third cumulant, etc. \cite{nazarov}. However, in order to fully characterise the device an analytical expression for the full counting statistics (FCS) is needed which provide information about correlations of all orders. They provide direct access to the probability distribution $P(Q)$ for the transfer of a certain amount of charge $Q$ during a certain waiting time $\tau$ via the cumulant generating function (CGF).\\
Considering the Anderson impurity model with a superconducting lead there are several energy scales one has to mind including the tunnel rate $\Gamma$, the onsite energy $U$ and the superconductor gap $\Delta$. From the theoretical side it is thus tempting to analyse a limiting case from the onset. Examples include the $U\rightarrow \infty$ limit which was analysed in \cite{PhysRevB.61.3555,PhysRevLett.80.2913} and the $\Delta\rightarrow\infty$ limit which was analysed via a rate equation approach in \cite{Braggio2011155}. These mappings to specific, again non-interacting, cases cannot show specific correlation effects. Furthermore, in actual experiments these limits can never be completely justified since typically the rates are of order $\Delta$ \cite{schoene2} and $U/\Delta$ can be around 6.5 \cite{PhysRevB.63.094515}.\\
The, so far only, approaches to the full complexity of the Anderson impurity model with a superconducting lead are numerical and presented in \cite{PhysRevB.63.094515,PhysRevB.84.075484} for the current and in \cite{PhysRevB.85.035419} for the cross correlation of currents. While the latter approach is limited to small interaction strength $U$ the former approach also allows for strong interaction by using an interpolative scheme for the electron self-energy which has a similar form in the weak and strong interaction case. For the case of weak interaction the authors of \cite{PhysRevB.63.094515,PhysRevB.84.075484} perform perturbation theory to second order in the Coulomb interaction. Since we aim at an analytical calculation of the CGF this calculation represents the first step \cite{PhysRevB.73.195301}. Our calculation has to be analytical since otherwise the relevant charge transfer processes cannot be extracted \cite{PhysRevLett.91.187001}.\\
We start by considering a single-impurity Anderson model written in terms of fermionic operators $d_\sigma^+, \; d_\sigma$ as
\begin{eqnarray}
H_D = \sum_\sigma \epsilon_D d_\sigma^+ d_\sigma,
\end{eqnarray}
where we use units such that $e = \hbar = k_B = 1$. The dot level $\epsilon_D$ can be experimentally varied by a gate electrode. We assume the usual hopping Hamiltonian between the dot and the fermionic leads \cite{PhysRevLett.8.316}
\begin{eqnarray}
H_T = \sum_\sigma \sum_\alpha \gamma [d_\sigma^+ \Psi_{\alpha \sigma}(x=0) + \mathrm{H.c.}],
\end{eqnarray}
where $\alpha = L,R = +,-$ refers to the two leads attached to the quantum dot and for simplicity we have taken a symmetric tunneling coupling from the onset. The normal lead $\alpha = R$ is described as a fermionic continuum by $H_R$ written in terms of creation and annihilation operators $\Psi_{R}$ at chemical potential $\mu_R$. The DOS $\rho_0$ is assumed to be constant so that the tunneling coupling $\gamma$ corresponds to a dot-lead tunnel rate of $\Gamma = \pi \rho_0 \gamma^2/2$. The superconductor is described in terms of the usual BCS Hamiltonian
\begin{eqnarray}
H_L &=& \sum_{\sigma} \epsilon_k \Psi_{L,k\sigma}^+ \Psi_{L,k\sigma} \nonumber\\
&& - \Delta \sum_k (\Psi_{L,k\uparrow}^+ \Psi_{L,-k\downarrow}^+ + \Psi_{L,-k\downarrow} \Psi_{L,k\uparrow}),
\end{eqnarray}
where we assume the superconducting gap $\Delta$ to be real and the quasiparticle DOS is given by $\rho_S = \rho_{0S} |\omega|/\sqrt{\omega^2 - \Delta^2}$. The corresponding dot-lead tunnel rate $\Gamma_S = \pi \rho_{0S} \gamma^2/2$ is assumed to be equal to the dot-lead tunnel rate $\Gamma$ on the normal conducting side.\\
In addition to these terms one has to take the electrostatic interaction into account
\begin{eqnarray}
H_U = U n_{D\uparrow} n_{D\downarrow} = U d_\uparrow^+ d_\uparrow d_\downarrow^+ d_\downarrow, \label{hint}
\end{eqnarray}
which reflects the energetic cost $U$ for double occupation. The full Hamiltonian is given by $H= H_R + H_L + H_T + H_D + H_U$.\\
The technology for the calculation of the FCS is by now far advanced. We calculate the CGF $\ln \chi(\lambda)$ as the fundamental quantity describing low-frequency transport \cite{PhysRevB.73.195301} and depending on the counting field $\lambda$. It has been shown \cite{nazarov} that $\ln \chi(\lambda)$ can be expressed in terms of Keldysh Green's function (GFs) so that the concept can be applied to numerous quantum impurity problems \cite{PhysRevB.83.085401}. The fundamental expression for calculating the CGF is \cite{levitov2004counting}
\begin{eqnarray}
\chi (\lambda) = \left \langle T_{\cal C} \exp \left[-i \int_{\cal C} dt H_T^{\lambda(t)} \right]\right\rangle_0,
\end{eqnarray}
where $T_{\cal C}$ denotes time-ordering along the Keldysh contour $\cal C$ and the expectation value is written in the interaction picture with respect to the Hamiltonian $H_L + H_R + H_D + H_U$. The dependence on the counting fields is contained in
\begin{eqnarray}
H_T^{\lambda(t)} &=& \sum_{\sigma} \gamma [e^{i \alpha \lambda/2} d_\sigma^+ \Psi_{R\sigma} + \mathrm{h.c.}] \nonumber\\
&& + \sum_{\sigma} \gamma [d_\sigma^+ \Psi_{L\sigma} + \mathrm{h.c.}] \label{htlambda}
\end{eqnarray}
The counting field is nonzero only during the (long) measurement time $\tau$ and has different signs on the forward and backward branch of the Keldysh contour.\\
Using the same technique as in \cite{PhysRevB.73.195301} the CGF may be written using an adiabatic potential $U_a(\lambda)$ (that by construction does not depend on time) \cite{PhysRevLett.26.1030} as
\begin{eqnarray}
\ln \chi(\lambda) = -i \tau U_a(\lambda).
\end{eqnarray}
Therefore the adiabatic potential is directly related to the counting field derivative of Eq. (\ref{htlambda}) via
\begin{eqnarray}
\frac{\partial U_a(\lambda)}{\partial \lambda^-} = - \frac{i \gamma}{2} \sum_{\sigma} \left[\langle T_{\cal C} \Psi_{R\sigma}^+ d_\sigma \rangle_\lambda e^{-i \alpha \lambda/2} + \mathrm{h.c.}\right]. \label{uaeq}
\end{eqnarray} 
We proceed by defining the exact-in-tunneling $\lambda$-dependent dot GF ${\cal D}_0^\lambda$ and the local (taken at $x=0$) free electrode GFs $g_\alpha$ in Keldysh space by \cite{jonckheere2009nonequilibrium}
\begin{eqnarray}
{\cal D}_{0\sigma}^\lambda(t,t') &=& -i \langle T_{\cal C} d_\sigma(t)d_\sigma^+(t') \rangle\\
g_{\alpha\sigma}(t,t') &=& -i \langle T_{\cal C} \Psi_{\alpha\sigma}(0,t) \Psi_{\alpha\sigma}^+(0,t') \rangle_0.
\end{eqnarray}
The expectation value $\langle \cdots \rangle$ is taken with respect to $H_L+ H_R + H_D + H_T^{\lambda(t)}$. Both spin-directions have the same GF so that we omit the additional spin index from now on.
In turn Eq. (\ref{uaeq}) can be written as the following convolution
\begin{eqnarray}
\frac{\partial U_a(\lambda)}{\partial \lambda^-} &=& - \frac{\gamma^2}{2} \int \frac{d\omega}{2\pi} \left[e^{i \lambda} g_{R}^{+-}(\omega) {\cal D}^{\lambda-+}(\omega) \right. \nonumber\\
&& \left. - e^{-i \lambda} g_{R}^{-+}(\omega) {\cal D}^{\lambda+-}(\omega) \right],
\end{eqnarray}
where upper indices indicate the Keldysh time contour of the two original time indices of the Keldysh GF. Mind that the above result only contains the normal components of the full dot GF and does not depend on the anomalous components of the dot GF even in presence of the SC lead.\\
As discussed before we will not perform perturbation theory in tunneling \cite{PhysRevB.77.134513} but perform perturbation theory in the interaction so that we can start from the non-interacting result $\ln \chi_0(\lambda)$ from \cite{soller2011charge}. There, we calculated the exact-in-tunneling GFs ${\cal D}_0^\lambda, {\cal C}_0^\lambda$ and ${\cal C}_0^{\lambda+}$ for the quantum dot and the corresponding CGF. The additional GFs are defined as ${\cal C}_{kk'}^{\lambda} = i \langle T_{\cal C} \Psi_{L,-k\downarrow}(t) \Psi_{L,k'\uparrow}(t')\rangle_\lambda$ and $\tilde{\cal C}_{kk'}^{\lambda} = i \langle T_{\cal C} \Psi_{L,k\uparrow}^+(t) \Psi_{L,-k'\downarrow}(t')\rangle_\lambda$.\\
Here we give the results for $T=0$ and in the limit of large bias $V\gg \Delta$ ($\ln \chi_{01}$) and small bias $V\ll \Delta$ ($\ln \chi_{02}$)
\begin{eqnarray}
\ln \chi_{01}(\lambda) &=& \tau \int_0^V \frac{d\omega}{\pi} \ln \left\{1+ \frac{4 \Gamma^2}{(\omega - \epsilon_D)^2 +4 \Gamma^2} \right. \nonumber\\
&& \times \left. (e^{i \lambda}-1)\right\}, \label{se0}\\
\ln \chi_{02}(\lambda) &=& \tau \int_{-V}^V \frac{d\omega}{2\pi} \ln \left\{1+ \frac{4 \Gamma^4}{[(\omega - \epsilon_D)^2 + 2 \Gamma^2]^2} \right. \nonumber\\
&& \left. \times (e^{2i \lambda} -1)\right\}, \label{ar0}
\end{eqnarray}
We observe the obvious difference between Eq. (\ref{se0}) and Eq. (\ref{ar0}) being the change of the exponential behavior on $\lambda$. The dependence in Eq. (\ref{se0}) refers to single-electron tunneling whereas the dependence on $e^{2i\lambda}$ in Eq. (\ref{ar0}) refers to double-electron transfer or Andreev reflection.\\
Additionally taking into account the interaction described by the Hamiltonian in Eq. (\ref{hint}) the full dot GF in Keldysh space is given by
\begin{eqnarray}
{\cal D}^\lambda(\omega) &=& {\cal D}_0^\lambda(\omega) + {\cal D}^{\lambda}(\omega) \Sigma_1(\omega) {\cal D}_0^\lambda(\omega) \nonumber\\
&& + {\cal D}^\lambda(\omega) \Sigma_2(\omega) \tilde{\cal C}_0^{\lambda}(\omega).
\end{eqnarray}
The self-energy diagrams to lowest order are illustrated in Fig. \ref{fig1}.
\\
$\Sigma_1$ corresponds to the self-energy caused by modifications to the dot GF as in Fig. \ref{fig1} (a), (b) and (c) and $\Sigma_2$ is due to diagrams as in Fig. \ref{fig1} (d), (e) and (f).\\
We calculate the second order contributions to the self-energy $\Sigma_1$ (Fig. \ref{fig1} (a), (b) and (c)) in the time domain
\begin{eqnarray*}
\Sigma_1 (t) &=& \left[ \begin{array}{l} \begin{array}{l} -i U\delta(t) {\cal D}_0^{\lambda--}(0) \\ + U^2 [{\cal D}_0^{\lambda--}(t)]^2 {\cal D}_0^{\lambda--}(-t) \\ + U^2 {\cal C}_0^{\lambda--}(t) \tilde{\cal C}_0^{\lambda--}(t) {\cal D}_0^{\lambda--}(-t) \\ \cr \end{array} \\ \begin{array}{c} - U^2 [{\cal D}_0^{\lambda+-}(t)]^2 {\cal D}_0^{\lambda -+}(-t) \\ - U^2 {\cal C}_0^{\lambda+-}(t) \tilde{\cal C}_0^{\lambda+-}(t) {\cal D}_0^{\lambda-+}(-t) \end{array} \end{array}\right. \nonumber\\
&& \left. \begin{array}{l} \begin{array}{c} -U^2 [{\cal D}_0^{\lambda-+}(t)]^2 {\cal D}_0^{\lambda+-}(-t) \\ - U^2 {\cal C}_0^{\lambda-+}(t) \tilde{\cal C}_0^{\lambda-+}(t) {\cal D}_0^{\lambda +-}(-t) \\ \cr \end{array} \\ \begin{array}{l} i U \delta(t) {\cal D}_0^{\lambda++}(0) \\ + U^2 [{\cal D}_0^{\lambda++}(t)]^2 {\cal D}_0^{\lambda++}(-t) \\ + U^2 {\cal C}_0^{\lambda++}(t) \tilde{\cal C}_0^{\lambda++}(t) {\cal D}_0^{\lambda++}(-t) \end{array} \end{array}\right].
\end{eqnarray*}
The linear part in $U$ is a remnant of the occupation probability of the dot level and can be evaluated as
\begin{eqnarray}
-i U {\cal D}_0^{\lambda--} (0) &=& -i U {\cal D}_0^{\lambda-+}(0) = -i U \int_{-\Delta}^\Delta \frac{d\omega}{2\pi} {\cal D}_0^{\lambda-+}(\omega)\nonumber\\
&=& Un_\lambda. \label{onefirst}
\end{eqnarray}
Using this result for $n_\lambda$ in Eq. (\ref{uaeq}) leads to an expression for the total CGF which is identical to the non-interacting result $\ln \chi_0(\lambda)$ presented in \cite{soller2011charge} up to the result for the denominator in the transmission coefficients where the bare level-energy is renormalised by $\tilde{\epsilon}_{D} \rightarrow \epsilon_{D} + Un_\lambda$. Subsequent expansion in $U$ and integration over energy results in a well controlled contribution which vanishes for the symmetric Anderson model $\epsilon_{D} =V/2 - U/2$ \cite{wiethege} to which case the following considerations are restricted.\\
Compared to the case of two normal conductors the superconducting gap represents an additional energy scale in the problem. Since the effect of all other contributions heavily depends on the ratio of the applied voltage $V$ and $\Delta$ analytical progress is limited to the cases $\Delta \ll V \ll \Gamma$ and $V\ll \Delta\ll \Gamma$.\\
The first case can be dealt with analogous to the treatment for normal conductors presented in \cite{PhysRevB.73.195301} since only $\Sigma_1$ will contribute to the self-energy. Adding all contributions which arise to second order in $U$ we can conclude that up to order $O(\Gamma^{-4})$ the correction to the CGF to be added to the non-interacting result in \cite{soller2011charge} for $V\gg \Delta$ is given by 
\begin{eqnarray}
&&\ln \chi_1^{(2)}(\lambda) =\nonumber\\
&& + \frac{4U^2 \tau V^3}{3 \pi^3 \Gamma^4} \left[ 2(e^{- 2i \lambda} -1) + (e^{-i \lambda }-1)\right] \nonumber\\
&& + \frac{8U^2 \tau V^3}{3 \pi^3 \Gamma^4} \left(3- \frac{\pi^2 }{4}\right) (e^{-i \lambda} -1). \label{corr12}
\end{eqnarray}
In this case we have no terms due to broken particle-hole symmetry as in Eq. (\ref{onefirst}) due to our specific choice of the dot level to be $\epsilon_D = V/2- U/2$. The first line is the most remarkable since it can be interpreted as correlated tunneling of electron pairs which consist of two electrons with opposite spin. This interpretation can be substantiated by considering the spin-dependent case \cite{PhysRevB.76.241307} or the Toulouse point of the corresponding Kondo Hamiltonian in the normal conducting case \cite{PhysRevB.73.195301}. Consequently, the onsite interaction leads to correlations similar to those encountered during Andreev reflection (AR).\\
The case $V\ll \Delta$ is the most relevant one since it shows clear signatures of SC correlations and we need to inspect both $\Sigma_1$ and $\Sigma_2$. The latter corresponds to the self-energies in Fig. \ref{fig1} (d), (e) and (f) and can be written in the time domain as
\begin{eqnarray*}
\Sigma_2(t) &=& \left[\begin{array}{c} \begin{array}{l} -i U\delta(t) {\cal C}_0^{\lambda+--}(0) \\ + U^2 [{\cal C}_0^{\lambda--}(t) \tilde{\cal C}_0^{\lambda--}(t)] {\cal C}_0^{\lambda--}(-t) \\ + U^2 [{\cal D}_0^{\lambda--}(t) {\cal C}_0^{\lambda--}(t)] {\cal D}_0^{\lambda--}(-t) \\ \cr \end{array} \\ \begin{array}{l} - U^2 [{\cal D}_0^{\lambda+-}(t) \tilde{\cal C}_0^{\lambda+-}(t)] {\cal C}_0^{\lambda-+}(-t) \\ - U^2 [{\cal C}_0^{\lambda+-}(t) \tilde{\cal C}_0^{\lambda+-}(t)] {\cal D}_0^{\lambda-+}(-t) \end{array} \end{array}\right. \nonumber\\
&& \left. \begin{array}{c} \begin{array}{l} - U^2 [{\cal D}_0^{\lambda-+}(t) {\cal C}_0^{\lambda-+} (t)] {\cal D}_0^{\lambda+-}(-t) \\ - U^2 [{\cal  C}_0^{\lambda-+}(t) \tilde{\cal C}_0^{\lambda-+}(t)] {\cal C}_0^{\lambda+-}(-t) \\ \cr \end{array} \\ \begin{array}{l} iU\delta(t) \tilde{\cal C}_0^{\lambda++}(t) \\ + U^2 [{\cal C}_0^{\lambda++}(t) \tilde{\cal C}_0^{\lambda++}(t)] {\cal C}_0^{\lambda++}(-t) \\ + U^2 [{\cal D}_0^{\lambda++}(t) {\cal C}_0^{\lambda++}(t)] {\cal D}_0^{\lambda++}(-t) \end{array} \end{array}\right].
\end{eqnarray*}
The linear part in $U$ is a remnant of the finite SC gap on the dot $\langle d_\uparrow^+ d_\downarrow^+\rangle$. As for the normal GF we can evaluate it via
\begin{eqnarray}
-i U \tilde{\cal C}_0^{\lambda--} (0) = -i U \int \frac{d\omega}{2\pi} \tilde{\cal C}_0^{\lambda--}(\omega).
\end{eqnarray}
The contribution we obtain is always finite and does not vanish for the symmetric Anderson model. However, insertion of this result into Eq. (\ref{uaeq}) and integration with respect to $\lambda$ leads to a well controlled contribution which is of the same form as the original CGF in \cite{soller2011charge}. Therefore, this contribution can be taken into account as an effective change of the local pairing potential $\Delta \rightarrow \Delta_{\mathrm{eff}}$ in the new CGF $\ln \tilde{\chi}_0(\lambda)$ as it has been done in previous works \cite{PhysRevB.68.035105}.\\
The quadratic part in $U$ is the most relevant one. As was done in \cite{PhysRevB.84.075484} for the current we evaluate the necessary susceptibilities for the self-energies $\Sigma_1$ and $\Sigma_2$, however, doing all necessary steps analytically and including the dependence on the counting field to obtain the expression for the FCS.\\
We integrate Eq. (\ref{uaeq}) with respect to $\lambda$ both for the terms containing $\Sigma_1$ and separately for those containing $\Sigma_2$. The prefactors are given as numbers as not all integrals have analytical solutions and all prefactors have been evaluated to leading order in $V$.\\
In total we obtain the following correction to second order in $U$ for the CGF
\begin{eqnarray}
\ln \chi_2^{(2)} &=& \frac{\tau U^2 V \Delta^2}{\pi^2 \Gamma^4} \left[(-0.001015) (e^{3i \lambda} -1) \right. \nonumber\\
&& + 0.017975 (e^{2i \lambda} -1) -0.00307 (e^{-2i \lambda} -1) \nonumber\\
&& - 0.00307 (e^{-i \lambda} -1) -0.002713 (e^{i\lambda} - 1) \nonumber\\
&& \left. + 0.0137 (e^{-3i \lambda} -1) + 0.0179 (e^{-4i \lambda} - 1)\right]. \nonumber\\
&& \label{fullcorrcgf}
\end{eqnarray}
We indeed observe correlation effects leading to a nontrivial dependence on $\lambda$. We want to discuss the terms according to their dependence on $\lambda$.\\
The processes in line two have the same dependence on $\lambda$ as typical Andreev reflections. These corrections are to be expected since the presence of Coulomb interaction on the dot leads to a different density of states on the dot which in turn has to affect the rate of Andreev processes.\\
The processes in line three arise due to the possibility of breaking a Cooper pair on the dot due to Coulomb repulsion. In a normal/inverse Andreev reflection two electrons would be transferred from/to the normal lead to/from the superconductor. However, Coulomb repulsion gives rise to breaking of Cooper pairs on the dot and the electrons can be transferred separately to the normal lead. Such breaking of Cooper pairs therefore gives rise to single-electron processes leading to terms $\propto e^{i \lambda}, e^{-i \lambda}$.\\
Finally the separated electrons on the dot do not have to enter the normal lead but can also be backscattered to the superconductor undergoing a second normal/inverse Andreev reflection. Such processes give rise to terms $\propto e^{3i\lambda}, e^{-3i\lambda}$.\\
Eq. (\ref{fullcorrcgf}) contains positive and negative exponents contrary to Eq. (\ref{corr12}) which contains only negative exponents. The difference is the effect of Andreev reflection. Whereas increased backscattering in case of normal leads only leads to reduced charge transport between the leads as can be seen from Eq. (\ref{corr12}) in case of a superconductor the electron on the dot may also undergo Andreev reflection and therefore charge transport can be enhanced. However, taking all possible charge transfer mechanisms in a normal-quantum dot-superconductor junction into account the effect of interaction is decreased conductance, as expected.
The most interesting term is the one $\propto e^{-4i \lambda}$. For voltages above the gap we have interpreted the occurence of double exponential terms as correlated tunneling of electron pairs. In this case single-electron tunneling to the SC is prohibited since charge transfer has to proceed via AR so that the correlated tunnel processes result in quartets in the CGF. This interpretation is substantiated by the fact that the final term in Eq. (\ref{fullcorrcgf}) is partially due to the purely normal part of the self-energy and therefore has the same origin as the correlated tunneling terms in Eq. (\ref{corr12}). Since the process of correlated tunneling of electron pairs in Eq. (\ref{corr12}) is a process of enhanced backscattering the same is true for the corresponding process in the superconducting case. The process leading to the quartic contribution is illustrated in Fig. \ref{fig33}.
\\
All terms in the CGF being linear in $V$ is a consequence of the anomalous dot GF which does not have the same clear cutoffs in energy at $T=0$ as the normal GF. All terms in Eq. (\ref{fullcorrcgf}) being linear in $V$ is reflected in the reduction of the zero bias conductance.\\
We can calculate the reduction of conductance from Eq. (\ref{fullcorrcgf}) and Eq. (\ref{ar0}) reintroducing SI-units to be
\begin{eqnarray}
\frac{G}{G_0} = 2 - 0.146596 \left(\frac{\Delta}{\Gamma}\right)^2  \left(\frac{U}{\Gamma}\right)^2. \label{condred}
\end{eqnarray}
We compared the resulting reduction of conductance to the one observed in \cite{PhysRevB.63.094515,PhysRevB.84.075484} which was also found to be proportional to $U^2$ for small interaction strenghth $U$ in accordance with our result. We also compared the two results on a quantitative level as shown in Fig. \ref{fig3}.
\\
We observe that our result correctly describes the qualitative behavior also for larger interaction strengths but fails to be quantitatively correct since we only performed second order perturbation theory. As expected the agreement is better for larger values of $\Gamma/\Delta$. However, since we performed perturbation theory in $U/\Delta$ for very large $\Gamma/\Delta$ the range of quantitative agreement in $U/\Gamma$ again becomes very small. Exact agreement of the two results should be expected additionally using the method of the interpolative self-energy as in \cite{PhysRevB.63.094515}.
The presence of strong current cross correlations found in \cite{PhysRevB.85.035419} could possibly be explained by presence of the additional correlated tunneling effects in Eq. (\ref{fullcorrcgf}).\\
However from Eq. (\ref{fullcorrcgf}) we know much more than the noise and the current. We know the full cumulant distribution function of charge which has been shown to be measurable in multiple experiments such as \cite{PhysRevB.74.195305}, where however, the interaction of the detector with the system in question was still sizeable.
Therefore the calculation of individual moments seems more rewarding. We can compute all moments from the CGF via $C_n = (-i)^n \partial^n/\partial \lambda^n [\ln \tilde{\chi}_0(\lambda) + \ln \chi_2^{(2)}(\lambda)]$ leading to universal values for $C_n/C_1$ for all $n$ and small voltage $V$ in accordance with previous results for normal conducting systems \cite{unitary}. These universal  values emerge as a consequence of the universal prefactor of the corrections in Eq. (\ref{fullcorrcgf}). Measuring the different cumulants therefore allows for a precise determination of the coefficients in CGF and the identification of the respective charge transfer processes.\\
These additional effects could be seen in experiments as the one suggested in \cite{PhysRevLett.98.056603}. There, the authors have calculated the cross correlation of currents for a quantum dot in the Kondo limit attached to three normal conducting leads and found a positive cross-correlation as an indication of the two-particle processes due to interaction that we have also observed in the $V\gg \Delta$ limit, see Eq. (\ref{corr12}). Likewise, we expect a positive fourth-order cross correlation of the currents in the normal leads in a structure with a superconductor connected to four normal conducting leads via a quantum dot as e.g. in \cite{soller2011charge}.\\
In conclusion we have calculated the first and second order corrections to the CGF for a normal-superconductor hybrid in an analytical manner to closely investigate the effects of onsite interaction on correlated charge transfer. For voltages above the gap we recovered the usual result known from normal conducting systems whereas for voltages below the gap we found a rich behavior. The most interesting result is the occurence of quartets in the CGF indicating correlated Andreev reflections through the quantum dot. Such processes give rise to four particle entanglement to be detected in futute experiments similar to those suggested in \cite{soller01}.

\end{document}